\documentclass[aps,prl,twocolumn,superscriptaddress,showpacs]{revtex4}
\usepackage{graphicx}
\usepackage{dcolumn}
\usepackage{bm}
\usepackage{color}
\usepackage{footmisc}
\usepackage{soul}

\begin{document}

\preprint{test}
\hyphenation{TiOCl}
\title{Two-Spinon and Orbital Excitations of the Spin-Peierls System TiOCl}

\author{S.~Glawion}
\author{J.~Heidler}
\affiliation{Experimentelle Physik 4, Universit\"at W\"urzburg, 97074 W\"urzburg, Germany}
\author{M.~W.~Haverkort}
\affiliation{Max Planck Institute for Solid State Research, 70569 Stuttgart, Germany}
\author{L.~C.~Duda}
\affiliation{Department of Physics and Materials Science, Uppsala Universitet, 751 21 Uppsala, Sweden}
\author{T.~Schmitt}
\author{V.~N.~Strocov}
\author{C.~Monney}
\author{K.~J.~Zhou}
\affiliation{Swiss Light Source, Paul Scherrer Institut, 5232 Villigen PSI, Switzerland}
\author{A.~Ruff}
\author{M.~Sing}
\author{R.~Claessen}
\affiliation{Experimentelle Physik 4, Universit\"at W\"urzburg, 97074 W\"urzburg, Germany}

\date{\today}

\begin{abstract}
We combine high-resolution resonant inelastic x-ray scattering with cluster
calculations utilizing a recently derived effective magnetic scattering operator
to analyze the polarization, excitation energy, and momentum-dependent
excitation spectrum of the low-dimensional quantum magnet TiOCl in the range
expected for orbital and magnetic excitations ($0-2.5$\,eV). Ti $3d$ orbital
excitations yield complete information on the temperature-dependent
crystal-field splitting. In the spin-Peierls phase we observe a dispersive
two-spinon excitation and estimate the inter- and intradimer magnetic exchange
coupling from a comparison to cluster calculations.
\end{abstract}

\pacs{78.70.Ck, 71.27.+a, 71.70.Ch, 78.70.Dm}

\maketitle

Understanding the interplay of lattice, charge, orbital, and spin degrees of
freedom in strongly correlated electron systems involving frustration remains
one of the great tasks in modern solid state physics. The role of electronic and
magnetic excitations for, e.g., the Mott metal-insulator transition is a key
question in this respect \cite{Imada98}. Resonant inelastic x-ray scattering
(RIXS) has been established recently as a powerful experimental technique to
study these phenomena in complex solids
\cite{Chiuz05,Kim07,Ulrich08,Ghiringhelli09,Braicovich09,Schlappa09,Ulrich09,
Ament09,Haverkort10,Braicovich10,Braicovich10b,Guarise10}, promoted by
considerably improved energy resolution and detector sensitivity
\cite{Ghiringhelli06,Strocov10}. While previously inelastic neutron scattering
was the method of choice for magnetic excitations, RIXS nowadays
is established to give complementary information already from much
smaller sample volumes ($< 0.1$\,mm$^3$). This is due to the much larger
scattering cross section of x rays compared to neutrons. Performing RIXS
with photon energies close to the $L$ absorption edge corresponds to processes
of the form $2p^6 3d^n \to 2p^5 3d^{n+1} \to 2p^6 3d^{n \ast}$. The intermediate
state is the final state of x-ray absorption spectroscopy (XAS), and the RIXS
final state can be an excited state (indicated by the asterisk) with losses due
to coupling to all kinds of excitations. For $n=1$ the initial and final state
can be described by single-particle wave functions, in contrast to the
intermediate state. Thus, although it is possible to extract the crystal-field
(CF) splitting directly from orbital ($dd$) excitations, a many-body approach is
necessary to simulate RIXS spectra.

In this Letter, we analyze the temperature, excitation energy, polarization, and
momentum dependence of the magnetic and orbital excitations of the
quasi-one-dimensional spin-Peierls system TiOCl measured by high-resolution
RIXS. The data allow us to extract precise values for the CF splitting in the room
temperature (RT) as well as the spin-Peierls phase. The polarization
dependence of the CF levels is in very good agreement with cluster calculations
based on density-functional theory within the local-density approximation.
From comparison to cluster calculations for a dimerized spin chain we identify a
low-energy feature as a two-spinon excitation, amplifying the
quasi-one-dimensional nature of TiOCl. The predictive power of a
recently derived effective scattering operator
for magnetic excitations in RIXS is evidenced for the first time by
simulating the intensity of this excitation for several different
momenta in relation to the $d$-$d$ excitation strength. Furthermore, by
including the dimerization into our theory such that the spin gap found in other
experiments \cite{Imai03} is reproduced we can infer the ratio of the inter- and
intradimer magnetic exchange coupling.

TiOCl is a layered system of buckled Ti-O bilayers sandwiched by Cl ions. Such
layers are stacked along the $c$ axis with only weak van der Waals forces
present in between, allowing for easy cleavage. The structure thus effectively
confines electronic and magnetic interactions to the $ab$ plane. In this plane,
Ti ions form a staggered triangular lattice which introduces magnetic
frustration. Locally, each Ti ion is coordinated by four O and two Cl ions which
form a strongly distorted octahedron with a local $C_{2v}$
point-group symmetry in the orthorhombic crystal structure. This results in a
completely lifted degeneracy of the $3d$ orbitals, i.e., five energetically well
separated CF levels, and thus a quenching of the orbital degrees of freedom
\cite{Saha-Dasgupta04}. The Ti ions are in a $3d^1$ configuration and the
overlap of occupied $d_{x^2 - y^2}$ orbitals along $b$ provides a one-dimensional
(1D) hopping path, leading to (quasi-)1D behavior in several aspects
\cite{Hoinkis05}. For example, below $T_{c1}=67$\,K the system shows a
dimerization of Ti ions along $b$ into $S=0$ singlets, i.e., a spin-Peierls
phase, with a huge spin gap of approximately 430\,K\,$= 37$\,meV \cite{Imai03}. At
higher temperatures the magnetic susceptibility corresponds to that of a 1D
Heisenberg chain, with a nearest-neighbor exchange coupling $J \approx
660$\,K\,$= 57$\,meV \cite{Seidel03}. The spin-Peierls ground-state below 67\,K is of
considerable interest since it is reached via an incommensurately modulated
intermediate phase below $T_{c2}=91$\,K \cite{Krimmel06}. Frustrated interchain
interactions are widely accepted as the driving force for this unconventional
behavior (cf. Ref.~\cite{Zhang08a} and references therein).

\begin{figure}
\includegraphics[width=0.48\textwidth]{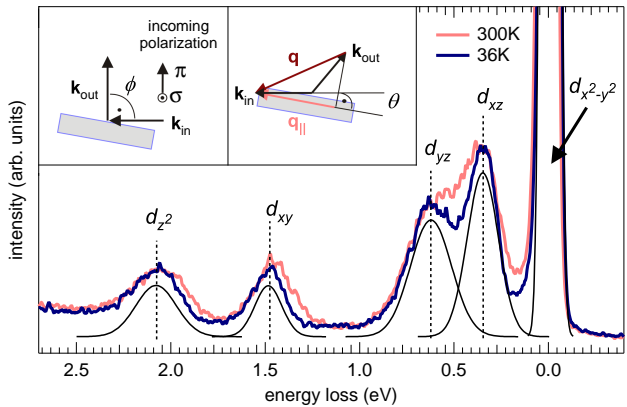}
\caption{\label{fig:ExpSetup}(Color online) RIXS spectra at $T=300$\,K
(light red) and $T=36$\,K (dark blue) ($h\nu =454.2$\,eV, $\sigma$ polarization,
$ac$ scattering; see text). Black solid curves are Gaussian fits to the low-$T$
data. Also, the orbital character of the CF excitations is indicated. Insets:
Experimental RIXS geometry. Left: $\phi=90^\circ$ geometry to define $\sigma$
(out-of-plane) and $\pi$ (in-plane) polarizations. Right: Definition of
$\boldsymbol{q}$ (brown arrow) and $q_{||}$ (red arrow). $\boldsymbol{q}$ was
fixed by the constant $\phi=50^\circ$ in our experiments, and $q_{||}$ was
varied by rotating the sample around the polar axis, i.e., changing $\theta$.}
\end{figure}

Samples were grown by a chemical vapor transport method \cite{Schaefer58}, with
a typical size of $(1\times 2\times 0.1)$\,mm$^3$. Measurements were conducted
at the SAXES endstation \cite{Ghiringhelli06,Strocov10} of the ADRESS beamline
(spot size $52\times4$\,$\mu$m$^2$) at the Swiss Light Source (Paul Scherrer
Institut, Switzerland) using soft x-ray RIXS with an energy resolution of
80\,meV or better, as determined from the full width at half maximum of the
elastic line of a carbon powder sample at the relevant excitation energies. As
shown in the insets of Fig.~\ref{fig:ExpSetup} the endstation allows different
angles $\phi$ between the incoming and outgoing beam, as well as a rotation of
the sample around the polar axis for the momentum-dependent measurements. For
$\phi=90^{\circ}$ and $\pi$-polarized (i.e., within the scattering plane)
incoming light the polarization of the outgoing beam must necessarily be
rotated. For $\sigma$-polarized (i.e., perpendicular to the scattering plane)
incoming light this need not be the case. This leads to well-defined
polarization effects even without detecting the polarization of the outgoing
photons. We used $\phi=90^{\circ}$ and a grazing incidence angle $\theta =
20^\circ$ for all RIXS measurements except those to investigate the momentum
dependence. In the latter case, only the projection of the momentum transfer
$\boldsymbol{q} = \boldsymbol{k}_{\text{in}} - \boldsymbol{k}_{\text{out}}$ on
the sample surface ($q_{||}$, see right inset of Fig.~\ref{fig:ExpSetup}) is of
importance due to the quasi-1D nature of TiOCl. Because of the bulk sensitivity of
RIXS governed by the attenuation length of the x rays ($>100$\,\AA), it was
sufficient to cleave samples in air using Scotch tape and then transfer them
into the vacuum ($<10^{-9}$\,mbar). Acquisition times per spectrum were in the
range $90-200$\,min. To simulate the $dd$ excitations we used a (paramagnetic)
TiO$_4$Cl$_2$ cluster within multiplet ligand-field theory, a powerful and
well-established many-body approach to x-ray absorption and emission processes,
especially in strongly correlated electron systems \cite{deGroot08}. Hopping
parameters and Slater integrals were obtained from local-density approximation calculations followed by
a downfolding procedure \cite{Haverkort11,Andersen00}. The cluster calculations
account for the CF splitting as well as the covalence between the Ti $3d$ and O
$2p$ orbitals. The magnetic excitations have been calculated from two separate
clusters utilizing a recently derived effective scattering operator for RIXS of
collective magnetic excitations \cite{Haverkort10}. The x-ray absorption spectra
that enter within this theory as the energy-dependent resonant enhancement
prefactors were obtained from the same cluster as the $dd$ excitations. For the
two-spinon excitations we used a cluster of 28 sites for a 1D dimerized Heisenberg
chain with nearest-neighbor exchange.

\begin{figure*}
\includegraphics[width=\textwidth]{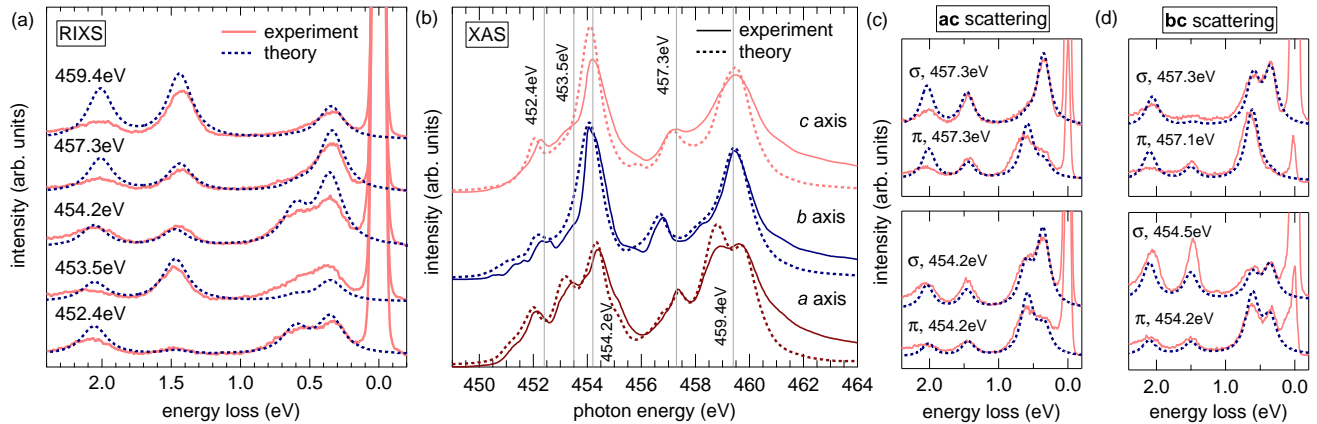}
\caption{\label{fig:PolDep}(Color online)(a) Excitation-energy dependence of
RIXS spectra at $T=300$\,K in $\sigma$ polarization and $ac$ scattering.
Corresponding excitation energies are indicated. (b) Ti $L$ edge XAS spectra
with main polarizations along the different crystal axes both measured (solid
curves) and calculated (dashed curves) for the geometry of panel (a). Photon
energies from (a) are marked by vertical lines. (c),(d) $\sigma$- and
$\pi$-polarized RIXS spectra for different photon energies in $ac$ (c) and $bc$
(d) scattering from experiments (solid curves) and multiplet calculations
(broken curves).}
\end{figure*}
Figure~\ref{fig:ExpSetup} shows typical RIXS spectra at $T=300$\,K (red curve)
and at $T=36$\,K (blue curve), i.e., in the spin-Peierls phase, taken at a
photon energy $h\nu=454.2$\,eV in $\sigma$ polarization with $ac$ being the
scattering plane (``$ac$ scattering''). In both cases, five peaks are
observable, corresponding to the on-site $dd$ excitations. For the first time,
all CF excitations of TiOCl are mapped directly in one experiment, as in other
(optical) probes some of them are forbidden due to dipole selection rules
\cite{Rueckamp05b}. Besides the elastic line at zero energy loss four peaks
appear at $0.34, 0.62, 1.48, \textrm{and } 2.08$\,eV \footnote{Error estimated
as $\pm0.01$\,eV.}, as determined from fitting Gaussian peaks to the low-$T$
data. At 300\,K, they lie at $0.36, 0.59, 1.45, \textrm{and } 2.05$\,eV,
respectively, due to $T$-dependent changes in the distortion of the
TiO$_4$Cl$_2$ octahedra predicted by cluster calculations \cite{Rueckamp05b}.

Figure~\ref{fig:PolDep}(a) shows RIXS spectra (solid red curves) at different
photon energies across the Ti $L$ edge ($T=300$\,K). Also shown are results from
cluster calculations (dashed blue curves), including 60\,meV resolution
and 200\,meV (final-state) lifetime broadening as inferred
from experiment. The elastic line has zero weight in the simulations
(except for Bragg reflections). This is expected for the perfectly ordered,
infinite crystal assumed in our calculations. In experiment, surface
reflectivity and scattering off crystal imperfections always produce a
(quasi-)elastic peak. The overall agreement between theory and experiment is
very good, and aside from sample quality the remaining discrepancies might be
connected to uncertainties concerning the exact excitation energy. This is
illustrated by Fig.~\ref{fig:PolDep}(b) which shows absorption spectra acquired
in total electron yield, with the (main) polarization \footnote{Due to the
experimental geometry all spectra have non-negligible admixtures from other
polarizations.} and the used photon energies indicated. The XAS consists of a
complicated multiplet structure, and being off in photon energy by a few meV can
lead to largely different intensities for the involved dipole transitions. Since
XAS is the intermediate state of RIXS the agreement between simulated and
measured RIXS spectra cannot be better than for the absorption spectra. Note
that the intermediate state broadening for the RIXS spectra
is equivalent to the broadening used for the calculation of the XAS spectra. The
core-hole broadening varies between 200\,meV for the low-energy states (450\,eV)
to 600\,meV for the high energy states (465\,eV). The
same also holds for the spectra shown in Figs.~\ref{fig:PolDep}(c) and (d),
where the effects of different polarization and scattering geometries are
compared at two excitation energies. Again, although some discrepancies remain
there is notable agreement between theory and experiment. Note that in the
experimental data the elastic line is strongly suppressed in $\pi$ polarization
\cite{Ghiringhelli04,vanVeenendaal06}.

\begin{figure}
\centering
\includegraphics[width=0.48\textwidth]{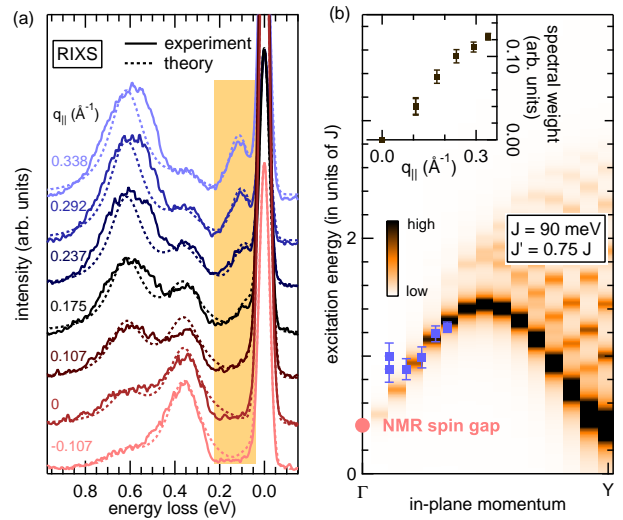}
\caption{\label{fig:qDep}(Color online)(a) RIXS spectra taken at
$h\nu=456.8$\,eV in $\pi$ polarization and $bc$ scattering ($T=15$\,K), for
$q_{||}$ corresponding to (from top to bottom) 36\%, 31\%, 25\%, 19\%, 11\%,
0\%, and -11\% of the $\Gamma$Y path in the BZ. The LEE is highlighted by the
shaded area. (b) Two-spinon dynamical spin structure factor for a 28 site
dimerized Heisenberg model. The dispersion of the LEE is indicated, obtained
from fitting the data in panel (a) (blue square markers). At $q_{||}=0$ the spin
gap extracted from NMR \cite{Imai03} is indicated. Inset: LEE spectral weight
(i.e., experimental peak areas).}
\end{figure}
A feature barely visible in the experimental spectra shown up to now is a
low-energy excitation (LEE) centered at roughly $0.1$\,eV, which is not captured
by the paramagnetic cluster calculations. Figure~\ref{fig:qDep}(a) shows
momentum-dependent RIXS spectra within an energy window containing only the two
lowest $dd$ excitations and the elastic line measured at $T=15$\,K for different
$q_{||}$ along $b$ ($\pi$ polarization, $bc$ scattering). Since
for some samples the LEE feature started to disappear after 5\,min of beam
exposure, the sample spot was changed every 5\,min by
10\,$\mu$m. Note, however, that also for such samples the behavior of the LEE
completely agrees with the data shown here, taken on samples, which did not
show any degradation of the spectra up to 20\,min of irradiation. Simulated
spectra in Fig.~\ref{fig:qDep}(a) consist of a superposition of three parts. In
addition to $dd$ excitations with a reduced lifetime broadening of 150 meV due
to the low measuring temperature, spinon calculations for the LEE based on the
above-mentioned effective operator \cite{Haverkort10}, and a Gaussian to mimic
the elastic peak are incorporated. Since at $h\nu=460$\,eV one has
$q\approx0.413$\,\AA$^{-1}$ and the Brillouin zone (BZ) boundary along the 1D direction (the Y
point) lies at 0.94\,\AA$^{-1}$ we could cover approximately $1/3$ of the path $\Gamma
\textrm{Y}$ with $q_{||}$.

The LEE is most pronounced in the spectrum with the highest $q_{||} =
0.338$\,\AA$^{-1}$ [top of Fig.~\ref{fig:qDep}(a)]. Towards the $\Gamma$ point
it rapidly loses intensity, and a dispersion from $\sim110$\,meV to
$\sim80$\,meV is observed. Its nature becomes evident from looking at the
two-spinon dynamical structure factor for a 1D chain
\cite{Mueller81,Karbach97}. It describes a continuum of excited states with two
sinelike (quasi-)boundaries $\omega_L(q)$ and $\omega_U(q)$, and its spectral
weight $\int^{\omega_U(q)}_{\omega_L(q)} S(q,\omega) \text{d}\omega$ goes
linearly to zero towards the zone center. To take dimerization into account we
calculated the dynamical structure factor for a 28 site Heisenberg spin chain with intra- and
interdimer nearest-neighbor interactions $J$ and $J'$, respectively. As can be
seen in Fig.~\ref{fig:qDep}(b) a spin gap opens at the zone center and the zone
boundary. The intensity, however, still goes linearly to zero towards the zone
center as for the undimerized chain. Note that in the experiment the in-plane
momentum and the polarization are changed simultaneously between different
spectra in Fig.~\ref{fig:qDep}(a) due to the experimental setup. This has been
accounted for in the calculations, but the dependence of the spectral weight on
the polarization for this specific case is only of the order of 10\%.
Consequently, there is good agreement for the intensity ratios of both the spin
and the orbital excitations. The experimental dispersion (blue square markers)
and the spin gap from NMR (red dot, $37$\,meV) \cite{Imai03} are best reproduced
for exchange couplings of $J=90$\,meV and $J'=0.75 J$, respectively, i.e., their
mean value is in reasonable agreement with the value of $57$\,meV for the
undimerized chain. The inset shows that the experimental LEE weight goes
linearly to zero, evidencing its two-spinon character. An analogous behavior of
magnetic features in RIXS has previously been observed e.g. on La$_2$CuO$_4$
\cite{Braicovich10b}, but TiOCl is the first noncuprate in which such an
excitation could be identified.

In principle, $dd$ excitations might also be dispersive. The orbital exchange
energy is of the same order as the magnetic exchange and smaller than their
lifetime broadening. Since in RIXS the probability for flipping an orbital or a
spin is roughly the same the experimental excitations likely involve coupled
orbital and magnon excitations, eventually with additional damping from phonons.
Looking back at Fig.~\ref{fig:qDep}(a) one notices that at large $q_{||}$, e.g.,
the $d_{yz}$ excitation is significantly broader than the local cluster
calculation. A detailed analysis of this observation, however, is out of the
scope of the present study.

In conclusion, by overcoming the selection-rule limitations of optical
spectroscopy using RIXS we have directly obtained reliable values for the CF
splitting from the orbital ($dd$) excitations of TiOCl. Their temperature,
excitation energy, and polarization dependence is in agreement with results from
cluster calculations. Evaluating the momentum-dependent behavior of a low-energy
peak observed in the spin-Peierls phase on the energy scale of the magnetic
exchange coupling $J$, we can identify this peak as a two-spinon excitation.
Using a new effective operator for magnetic excitations in RIXS in
combination with finite size cluster calculations its dispersion and
intensity is accounted for. Moreover, we find robust values for
the inter- and intra-dimer exchange coupling.

\begin{acknowledgments}
We acknowledge fruitful discussions with F. F. Assaad and J. Schlappa. This work
was supported by the German Science Foundation (DFG) via Grants No. CL-124/6-1
and No. CL-124/6-2, the Swedish Research Council (VR), and the G\"oran Gustafsson
Foundation (GGS). This work was performed at the ADRESS beamline using the SAXES
instrument jointly built by Paul Scherrer Institut, Switzerland and Politecnico
di Milano, Italy.

\end{acknowledgments}


\end{document}